\RequirePackage{fix-cm}
\documentclass[smallextended]{svjour3}       % onecolumn (second format)
\smartqed  % flush right qed marks, e.g. at end of proof
\usepackage{graphicx}
\usepackage{amsmath}
\usepackage{amssymb}
\begin{document}

\title{Lepton flavor violation and seesaw symmetries}

\author{D. Aristizabal Sierra}

\institute{Diego Aristizabal Sierra \at IFPA,
  Department AGO, Universite de Liege, Bat
  B5 \small \sl Sart Tilman B-4000 Liege 1, Belgium.\\
  \email{daristizabal@ulg.ac.be}} 
\date{}

\maketitle

\begin{abstract}
  When the standard model is extended with right-handed neutrinos the
  symmetries of the resulting Lagrangian are enlarged with a new
  global $U(1)_R$ Abelian factor. In the context of minimal seesaw
  models we analyze the implications of a slightly broken $U(1)_R$
  symmetry on charged lepton flavor violating decays. We find,
  depending on the $R$-charge assignments, models where charged lepton
  flavor violating rates can be within measurable ranges. In
  particular, we show that in the resulting models due to the
  structure of the light neutrino mass matrix muon flavor violating
  decays are entirely determined by neutrino data (up to a
  normalization factor) and can be sizable in a wide right-handed
  neutrino mass range.
  
\end{abstract}

\section{Introduction}
\label{sec:intro}
Apart from demonstrating that neutrinos are massive and have
non-vanishing mixing angles among the different generations
\cite{Schwetz:2011zk,concha}, neutrino oscillation experiments have
also proved that lepton flavor is not conserved in the neutral lepton
sector. Once the standard model is extended to account for neutrino
masses---unavoidably---lepton flavor violation (LFV) also takes place
in the charged lepton sector. This, however, not necessarily implies
that these effects are sizable, so whether these processes can or not
have measurable rates depends to a large extent on the details of the
corresponding model. Despite this fact, from a general point of view,
charged lepton flavor violating processes are expected to have large
decay branching fractions as long as the LFV mediators have ${\cal
  O}(\mbox{TeV})$ masses and their couplings to the standard model
leptons are about $\gtrsim 10^{-2}$.

Majorana neutrino masses can be generated in a model independent way
by adding the dimension-five effective operator ${\cal O}_5\sim LLHH$
to the standard model Lagrangian \cite{Weinberg:1980bf}. And in turn
the different concrete realizations of this operator constitute a
model for neutrino masses \footnote{Examples range from tree-level up
  to three-loop induced neutrino mass models
  \cite{Zee:1980ai,AristizabalSierra:2006ri,Zee:1985id,Babu:1988ki,Babu:2002uu,AristizabalSierra:2006gb,Nebot:2007bc,AristizabalSierra:2007nf,FileviezPerez:2009ud,Babu:2010vp}}. Among
the tree-level realizations the type-I seesaw is certainly the most
popular one
\cite{Minkowski:1977sc,Mohapatra:1979ia,Yanagida:1979as,GellMann:1980vs,Glashow:1979nm,Schechter:1980gr}. In
this model, light neutrino masses are generated via the exchange of
electroweak fermionic singlets (right-handed (RH) neutrinos for
brevity). Consistency with neutrino data then requires either heavy RH
neutrino masses (${\cal O}(M_N)\sim \Lambda_\text{GUT}$) or tiny
Yukawa couplings (${\cal O}\sim 10^{-6}$), thus implying negligibly
small charged lepton flavor violating effects.

In addition to the standard model gauge symmetry the seesaw Lagrangian
features a global Abelian $U(1)_R$ symmetry, typically related with
phase rotations of the standard model lepton $SU(2)$ singlets, and
thus broken by the charged lepton Yukawa couplings. However, relating
this symmetry with phase rotations of the RH leptons fields is not the
only possibility, and another approach in which rotations of the
left-handed lepton fields and RH neutrinos are allowed is feasible as
well. In that case one is left with (at least) two choices: ($i$)
slightly broken $U(1)_R$; ($ii$) $\mathbb{Z}_n\subset U(1)_R$
invariance of the Lagrangian.

In what follows we will consider possibility ($i$). In the context of
minimal seesaw models (models featuring only 2 RH neutrinos) we will
classify the viable scenarios arising from different $R$-charge
assignments, that as we already discussed are not anymore limited to
the RH leptons, and identify those models for which charged lepton
flavor violating processes have sizable decay branching ratios. For
these models we will analyze the $\mu$ flavor violating
phenomenology. The results presented here are entirely based on
ref. \cite{AristizabalSierra:2012yy}.
\section{The models}
\label{sec:models}
Depending on the $R$-charge assignments of the different standard
model and RH neutrino fields different models can be constructed. In
order to restrict the discussion only to the lepton sector we start by
setting $R(H)=0$. Requiring the charged lepton Yukawa couplings to be
$U(1)_R$ invariant allows to fix $R(e)=R(\ell)$ ($e,\ell$ being the
lepton electroweak singlets and doublets). We are thus left with the
lepton doublets and RH neutrinos $R$-charge assignments. Large lepton
flavor violating rates require (at least) the RH neutrino mass terms
to be $U(1)_R$ breaking suppressed (the suppression factor denoted by
$\epsilon\ll 1$), implying $R(N_{1,2})\neq 0$ and one of the following
three possibilities: (A) $R(N_1)=R(N_2)$; (B) $R(N_1)=-R(N_2)$ or (C)
$R(N_1)\neq R(N_2)$. In practice possibilities (A) and (C) turn out to
be equivalent as they both lead to models with $N_1-N_2$ suppressed
mixing and therefore to suppressed LFV effects. In contrast in case
(B) the $N_1-N_2$ mixing is maximal and a set of Yukawa couplings can
be large provided the $R(\ell)$ charges are chosen appropriately. In
that sense models of type (B) are much more interesting as they might
yield large LFV effects.

With the purpose of studying the implications for LFV of type B models
we fix the $R$-charges as $R(N_1,\ell_i,e_i)=+1$ and $R(N_2)=-1$. With
this charge assignment the seesaw Lagrangian becomes
\footnote{Phenomenologically these models are similar to models where
  lepton number is slightly broken (see e.g.  references
  \cite{Gavela:2009cd,Mohapatra:1986bd,branco-lavoura-grimus,abada-bonnet-gavela-hambye,Gu:2008yj,Ibanez:2009du,forero-tortoal-morisi-valle})}
\begin{equation}
  \label{eq:seesaw-lag-N!N2mismatch}
  {\cal L}=
  %-\bar \ell\,\pmb{\hat Y_e}\,e H
  - \bar \ell\,\pmb{\lambda_1}^*\,N_1 \tilde H
  - \epsilon_\lambda\, \bar \ell\,\pmb{\lambda_2}^*\,N_2 \tilde H
  - \frac{1}{2} N_1^T\,C M\,N_2
  - \frac{1}{2}\epsilon_N N_a^T\,C M_{aa}\,N_a + 
  \mbox{h.c.}\,.
\end{equation}
Here $\tilde H = i\sigma_2 H^*$, $C$ is the charge conjugation
operator,
$\pmb{\lambda_a}^\dagger=(\lambda_{1a}^*,\lambda_{2a}^*,\lambda_{3a}^*)$
with $a=1,2$ (matrices are denoted in bold-face) and
$\epsilon_{\lambda,N}$ are dimensionless parameters that slightly
break $U(1)_R$.  Diagonalization of the RH neutrino mass matrix yields
two quasi-degenerate states with masses
\begin{equation}
  \label{eq:RHN-masses-quasideg}
  M_{N_{1,2}}=M\mp\frac{M_{11}+M_{22}}{2}\epsilon_N\,.
\end{equation}
After diagonalization the Yukawa couplings read
\begin{equation}
  \label{eq:Yukawa-rotated}
  \lambda_{ka}\to -\frac{(i)^a}{\sqrt{2}}\left[\lambda_{k1}
    + (-1)^a\epsilon_\lambda\lambda_{k2}\right]\,,
  \qquad (k=e,\mu,\tau\;\;\mbox{and}\;\; a=1,2)\,,
\end{equation}
and thus the light neutrino mass matrix is determined to be
\begin{equation}
  \label{eq:light-nmm-N1N2mismatch}
  \pmb{m_\nu^\text{eff}}=-\frac{v^2\epsilon_\lambda}{M}
  |\pmb{\lambda_1}||\pmb{\Lambda}|
  \left(
    \pmb{\hat \lambda_1}^*\otimes\pmb{\hat \Lambda}^* 
    +
    \pmb{\hat \Lambda}^*\otimes\pmb{\hat \lambda_1}^*
  \right)\,,
\end{equation}
with
\begin{equation}
  \label{eq:Lambda}
  \pmb{\hat \Lambda}^*=\pmb{\hat \lambda_2}^*-
  \frac{M_{11}+M_{22}}{4M}\frac{\epsilon_\lambda}{\epsilon_N}
  \pmb{\hat \lambda_1}^*\,.
\end{equation}
Note that the parameter space vectors have been expressed according to
$\pmb{\lambda_1}=|\pmb{\lambda_1}|\pmb{\hat \lambda_1}$,
$\pmb{\Lambda}=|\pmb{\Lambda}|\pmb{\hat \Lambda}$, where $\pmb{\hat
  \lambda_1}, \pmb{\hat \Lambda}$ are unitary vectors along the
$\pmb{\lambda_1}, \pmb{\Lambda}$ directions. Due to the structure of
the light neutrino matrix the parameter space vectors are---up to
normalization factors---completely determined by neutrino mixing
angles and masses. For the normal hierarchical spectrum they can be
written according to \cite{Gavela:2009cd}:
  \begin{align}
    \label{eq:Yukawas-normalS1}
    \pmb{\lambda_1}&=|\pmb{\lambda_1}|\;\pmb{\hat \lambda_1}=
    \frac{|\pmb{\lambda_1}|}{\sqrt{2}}
    \left(
      \sqrt{1+\rho}\,\pmb{U_3}^* + \sqrt{1-\rho}\,\pmb{U_2}^*
    \right)\,,\\
    \label{eq:Yukawas-normalS2}
    \pmb{\Lambda}&=|\pmb{\Lambda}|\;\pmb{\hat \Lambda}=
    \frac{|\pmb{\Lambda}|}{\sqrt{2}}
    \left(
      \sqrt{1+\rho}\,\pmb{U_3}^* - \sqrt{1-\rho}\,\pmb{U_2}^*
    \right)\,,
  \end{align}
  where the $\pmb{U_i}$'s correspond to the columns of the leptonic
  mixing matrix and
\begin{equation}
  \label{eq:rho-and-r-NS}
  \rho=\frac{\sqrt{1+r}-\sqrt{r}}{\sqrt{1+r}+\sqrt{r}}\,,\qquad
  r=\frac{m_{\nu_2}^2}{m_{\nu_3}^2-m_{\nu_2}^2}\,.
\end{equation}
\section{Charged lepton flavor violating decays}
\label{sec:2}
Currently the most competitive bounds on charged lepton flavor
violating processes are placed for $\mu$ decays, being $\mu\to
e\gamma$, $\mu\to 3e$ and $\mu-e$ conversion in nuclei the ones with
the most stringent upper limits \cite{PDG}. In addition it is for these
processes that the most tight bounds are expected in near-future
experimental proposals: MEG \cite{meg-futureS}, {\it Mu3e} \cite{psi}
PRISM/PRIME \cite{Ankenbrandt:2006zu}.  So henceforth
we will focus on $\mu$ decays, in particular on the reactions $\mu\to
e\gamma$ and $\mu\to 3 e$ (for $\mu-e$ conversion in nuclei see
ref. \cite{AristizabalSierra:2012yy}).
\begin{figure}
  \centering
  \includegraphics[width=7cm]{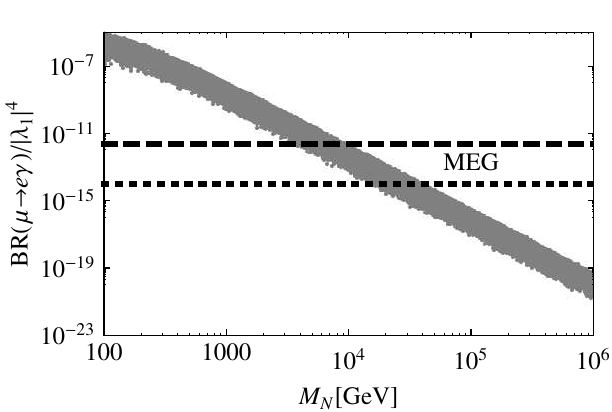}
  \caption{Decay branching ratio BR$(\mu\to e\gamma)$ normalized
    to $|\pmb{\lambda_1}|^4$ for the normal light neutrino mass
    spectrum as a function of the common RH neutrino mass. The upper
    horizontal dashed line indicates the current experimental upper
    limit from the MEG experiment ~\cite{Adam:2011ch}, whereas the
    lower dotted one marks prospective future experimental
    sensitivities~\cite{meg-futureS}.}
  \label{fig:radiative-lfv}
\end{figure}
\subsection{$\mu\to e \gamma$ process}
\label{sec:mutoegama}
In the limit $M_W/M\ll 1$ the decay branching ratio for this
decay can be written as \cite{Ilakovac:1994kj}
\begin{equation}
  \label{eq:4}
  \text{BR}(\mu\to e \gamma)\simeq\frac{\alpha}{1024 \pi^4}
  \frac{m_\mu^5}{M^4}\frac{|\pmb{\lambda_1}|^4}{\Gamma_\text{Tot}^{\mu}}
  \left|
    \hat\lambda_{21}\;\hat\lambda_{11}^*
  \right|^2\,.
\end{equation}
Thus showing that apart from the parameters $M$ and
$\pmb{|\lambda_1|}$ this branching fraction is entirely determined by
low-energy data (see eqs. (\ref{eq:Yukawas-normalS1}),
(\ref{eq:Yukawas-normalS2}) and (\ref{eq:rho-and-r-NS})). Figure
\ref{fig:radiative-lfv} shows the results obtained from the full
formula involving the complete one-loop function (see
ref. \cite{AristizabalSierra:2012yy} for details) and by randomly
generating the low-energy observables in their $2\sigma$ allowed range
\cite{Schwetz:2011zk,concha} (normal hierarchical mass spectrum), the
parameters $|\pmb{\lambda_1}|$ and $M$ in the intervals $[10^{-5},1]$
and $[10^2,10^6]$ GeV and the $N_{1,2}$ mass splittings in the range
$[10^{-8},10^{-6}]$ GeV. As can be realized from
eqs. (\ref{eq:Yukawas-normalS1}), (\ref{eq:Yukawas-normalS2}),
(\ref{eq:rho-and-r-NS}) and (\ref{eq:4}) the width of the band is due
to neutrino data uncertainties.

From fig. \ref{fig:radiative-lfv} it can be seen that BR$(\mu\to
e\gamma)$ can reach the current experimental upper bound
\cite{Adam:2011ch} as long as
$M_N<0.1~\mathrm{TeV},1~\mathrm{TeV},~10$~TeV provided
$|{\pmb{\lambda_1}}|\gtrsim 2\times 10^{-2},~10^{-1}, ~1$,
respectively.
\subsection{$\mu\to 3 e $ process}
\label{sec:mutoegama}
\begin{figure}
  \centering
  \includegraphics[width=7cm]{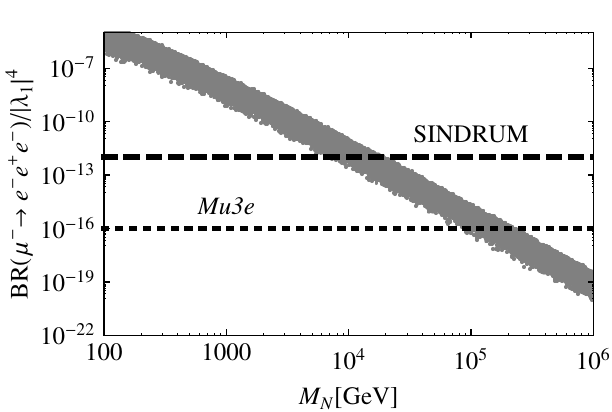}
  \caption{Decay branching ratio BR($\mu^-\to e^- e^+ e^-)$ normalized
    to $|\pmb{\lambda_1}|^4$ for normal light neutrino mass spectrum
    as a function of the common RH neutrino mass. The upper horizontal
    dashed line indicates the current bound on the $\mu^-\to
    e^+e^-e^-$ rate placed by the SINDRUM experiment
    ~\cite{Bellgardt:1987du}, whereas the lower dotted one illustrates
    prospective future experimental sensitivities of the {\it Mu3e}
    experiment~\cite{psi}.}
  \label{fig:muto3elec}
\end{figure}

We now turn to the discussion of the $\mu\to 3 e$ process. This decay
involves dipole contributions, $\gamma$ and $Z$ penguins as well as
box diagrams \cite{Ilakovac:1994kj}, so a simple approximate formula
as in the previous case does not exist. Following the same numerical
procedure than in the $\mu\to e\gamma$ case we calculate the
corresponding decay branching ratio. Fig. \ref{fig:muto3elec} shows
the result for the branching fraction normalized to
$|\pmb{\lambda_1}|^4$ for the normal hierarchical mass spectrum.

It can be seen that $\mbox{BR}(\mu\to 3e)$ can exceed the experimental
upper limit for RH neutrino masses
$M_N<0.1~\mathrm{TeV},1~\mathrm{TeV},~10$~TeV provided
$|{\pmb{\lambda_1}}|\gtrsim 2\times 10^{-2},~10^{-1}, ~1$,
respectively, very similar to the $\mu \to e \gamma$ case.  Mainly due
to the sensitivities of the planned future experiments
($10^{-16}-10^{-15}$) this decay has the potential to probe
considerably larger values of the RH neutrino masses (compared with
$\mu\to e\gamma$), reaching RH neutrino mass scales in excess of
$\mathcal O(10^5~{\rm GeV})$ for $|\pmb{\lambda_1}|\sim 1$. Finally we
note that due to the strong $|\pmb{\lambda_1}|$ dependence, values of
$|\pmb{\lambda_1}|$ below $10^{-3}$ are not expected to yield
observable rates at near future experimental facilities even for RH
neutrino masses of the order 100~GeV.

\section{Conclusions}
\label{sec:conclusions}
We studied the implications of the seesaw global Abelian $U(1)_R$
symmetry on lepton flavor violation, in the context of minimal seesaw
models. We showed that depending on the $R$-charge
assignments---generically---two type of models can be identified. A
first class where the mechanism that suppresses the light neutrino
masses propagates to the lepton flavor violating observables, thus
implying negligibly small LFV effects. A second class in which the
mechanism ``decouples'' yielding in that way sizable rates for lepton
flavor violating $\mu$ decays. We discussed $\mu\to e\gamma$ and
$\mu\to 3e$ and showed that these processes might have decay branching
fractions within the reach of current or near-future experiments.
\begin{acknowledgements}
  I want to thank Audrey Degee and Jernej F. Kamenik for the fruitful
  collaboration that led to the paper on which this article is based.
\end{acknowledgements}


\begin{thebibliography}{}
\bibitem{Schwetz:2011zk} T.~Schwetz, M.~Tortola and J.~W.~F.~Valle,
  ``Where we are on $\theta_{13}$: addendum to 'Global neutrino data
  and recent reactor fluxes: status of three-flavour oscillation
  parameters' '', New J.\ Phys.\ {\bf 13}, 109401 (2011)
  [arXiv:1108.1376 [hep-ph]];
  %%CITATION = ARXIV:1108.1376;%%

\bibitem{concha}
  M.~C.~Gonzalez-Garcia, M.~Maltoni and J.~Salvado, ``Updated global
  fit to three neutrino mixing: status of the hints of $\theta_{13}>0$'',
  JHEP {\bf 1004}, 056 (2010) [arXiv:1001.4524 [hep-ph]].
  %%CITATION = ARXIV:1001.4524;%%

\bibitem{Weinberg:1980bf}
  S.~Weinberg,
  %Varieties of Baryon and Lepton Nonconservation,
  Phys.\ Rev.\  D {\bf 22}, 1694 (1980).

%\cite{Zee:1980ai}
\bibitem{Zee:1980ai}
  A.~Zee,
  ``A Theory of Lepton Number Violation, Neutrino Majorana Mass, and Oscillation'',
  Phys.\ Lett.\  {\bf B93}, 389 (1980).

%\cite{AristizabalSierra:2006ri}
\bibitem{AristizabalSierra:2006ri} 
  D.~Aristizabal Sierra and D.~Restrepo,
  ``Leptonic Charged Higgs Decays in the Zee Model,''
  JHEP {\bf 0608}, 036 (2006)
  [hep-ph/0604012].
  %%CITATION = HEP-PH/0604012;%%

\bibitem{Zee:1985id}
  A.~Zee,
  ``Quantum Numbers Of Majorana Neutrino Masses'',
  Nucl.\ Phys.\  B {\bf 264}, 99 (1986).

\bibitem{Babu:1988ki}
  K.~S.~Babu,
  ``Model of 'Calculable' Majorana Neutrino Masses'',
  Phys.\ Lett.\  B {\bf 203}, 132 (1988).

%\cite{Babu:2002uu}
\bibitem{Babu:2002uu}
  K.~S.~Babu and C.~Macesanu,
  Two-loop neutrino mass generation and its experimental consequences,
  Phys.\ Rev.\  {\bf D67}, 073010 (2003) [hep-ph/0212058].
  %%Phys.\ Rev.\ D {\bf 67} (2003) 073010
  %[arXiv:hep-ph/0212058].
  %%CITATION = HEP-PH 0212058;%%

%\cite{AristizabalSierra:2006gb}
\bibitem{AristizabalSierra:2006gb}
  D.~Aristizabal Sierra, M.~Hirsch,
  ``Experimental tests for the Babu-Zee two-loop model of Majorana neutrino masses'',
  JHEP {\bf 0612}, 052 (2006).
  [hep-ph/0609307].

%\cite{Nebot:2007bc}
\bibitem{Nebot:2007bc}
  M.~Nebot, J.~F.~Oliver, D.~Palao, A.~Santamaria,
  ``Prospects for the Zee-Babu Model at the CERN LHC and low energy experiments'',
  Phys.\ Rev.\  {\bf D77}, 093013 (2008).
  [arXiv:0711.0483 [hep-ph]].

%\cite{AristizabalSierra:2007nf}
\bibitem{AristizabalSierra:2007nf}
  D.~Aristizabal Sierra, M.~Hirsch, S.~G.~Kovalenko,
  ``Leptoquarks: Neutrino masses and accelerator phenomenology'',
  Phys.\ Rev.\  {\bf D77}, 055011 (2008).
  [arXiv:0710.5699 [hep-ph]].

\bibitem{FileviezPerez:2009ud}
  P.~Fileviez Perez, M.~B.~Wise,
  ``On the Origin of Neutrino Masses'',
  Phys.\ Rev.\  {\bf D80}, 053006 (2009).
  [arXiv:0906.2950 [hep-ph]].

%\cite{Babu:2010vp}
\bibitem{Babu:2010vp}
  K.~S.~Babu, J.~Julio,
  ``Two-Loop Neutrino Mass Generation through Leptoquarks'',
  Nucl.\ Phys.\  {\bf B841}, 130-156 (2010).
  [arXiv:1006.1092 [hep-ph]].
% --------------------
% Seesaw references
% --------------------

%\cite{Minkowski:1977sc}
\bibitem{Minkowski:1977sc} 
  P.~Minkowski,
  ``$\mu \to e \gamma$ at a rate of one out of 1-Billion muon decays?'',
  Phys.\ Lett.\ B {\bf 67}, 421 (1977).
  %%CITATION = PHLTA,B67,421;%%

%\cite{Mohapatra:1979ia}
\bibitem{Mohapatra:1979ia} 
  R.~N.~Mohapatra and G.~Senjanovic,
  ``Neutrino Mass and Spontaneous Parity Violation'',
  Phys.\ Rev.\ Lett.\  {\bf 44}, 912 (1980).
  %%CITATION = PRLTA,44,912;%%

%\cite{Yanagida:1979as}
\bibitem{Yanagida:1979as} 
  T.~Yanagida,
  ``Horizontal Symmetry And Masses Of Neutrinos'',
  Conf.\ Proc.\ C {\bf 7902131}, 95 (1979).
  %%CITATION = CONFP,C7902131,95;%%

%\cite{GellMann:1980vs}
\bibitem{GellMann:1980vs} 
  M.~Gell-Mann, P.~Ramond and R.~Slansky,
  ``Complex Spinors And Unified Theories'',
  Conf.\ Proc.\ C {\bf 790927}, 315 (1979).
  %%CITATION = CONFP,C790927,315;%%

%\cite{Glashow:1979nm}
\bibitem{Glashow:1979nm} 
  S.~L.~Glashow,
  ``The Future Of Elementary Particle Physics'',
  NATO Adv.\ Study Inst.\ Ser.\ B Phys.\  {\bf 59}, 687 (1980).
  %%CITATION = NASBD,59,687;%%

%\cite{Schechter:1980gr}
\bibitem{Schechter:1980gr} 
  J.~Schechter and J.~W.~F.~Valle,
  ``Neutrino Masses in $SU(2) \times U(1)$ Theories'',
  Phys.\ Rev.\ D {\bf 22}, 2227 (1980).
  %%CITATION = PHRVA,D22,2227;%%

% --------------------------

%\cite{Alonso:2011jd}
\bibitem{Alonso:2011jd} 
  R.~Alonso, G.~Isidori, L.~Merlo, L.~A.~Munoz and E.~Nardi,
  %``Minimal flavour violation extensions of the seesaw,''
  JHEP {\bf 1106}, 037 (2011)
  [arXiv:1103.5461 [hep-ph]].
  %%CITATION = ARXIV:1103.5461;%%

%\cite{AristizabalSierra:2012yy}
\bibitem{AristizabalSierra:2012yy} 
  D.~Aristizabal Sierra, A.~Degee and J.~F.~Kamenik,
  %``Minimal lepton flavor violating realizations of minimal seesaw models,''
  JHEP {\bf 1207}, 135 (2012)
  [arXiv:1205.5547 [hep-ph]].
  %%CITATION = ARXIV:1205.5547;%%

%\cite{Gavela:2009cd}
\bibitem{Gavela:2009cd} 
  M.~B.~Gavela, T.~Hambye, D.~Hernandez and P.~Hernandez,
  ``Minimal Flavour Seesaw Models'',
  JHEP {\bf 0909}, 038 (2009)
  [arXiv:0906.1461 [hep-ph]].
  %%CITATION = ARXIV:0906.1461;%%

%\cite{Mohapatra:1986bd}
\bibitem{Mohapatra:1986bd}
  R.~N.~Mohapatra and J.~W.~F.~Valle,
  ``Neutrino Mass and Baryon Number Nonconservation in Superstring Models'',
  Phys.\ Rev.\ D {\bf 34}, 1642 (1986);
  %%CITATION = PHRVA,D34,1642;%%

\bibitem{branco-lavoura-grimus}
  G.~C.~Branco, W.~Grimus and L.~Lavoura,
  ``The Seesaw Mechanism In The Presence Of A Conserved Lepton Number'',
  Nucl.\ Phys.\ B {\bf 312}, 492 (1989);
  %%CITATION = NUPHA,B312,492;%%

\bibitem{abada-bonnet-gavela-hambye}
  A.~Abada, C.~Biggio, F.~Bonnet, M.~B.~Gavela and T.~Hambye,
  ``Low energy effects of neutrino masses'',
  JHEP {\bf 0712}, 061 (2007)
  [arXiv:0707.4058 [hep-ph]].
  %% CITATION = ARXIV:0707.4058;%%

\bibitem{Gu:2008yj} 
  P.~-H.~Gu, M.~Hirsch, U.~Sarkar and J.~W.~F.~Valle,
  ``Neutrino masses, leptogenesis and dark matter in hybrid seesaw'',
  Phys.\ Rev.\ D {\bf 79}, 033010 (2009)
  [arXiv:0811.0953 [hep-ph]].
  %%CITATION = ARXIV:0811.0953;%%

%\cite{Ibanez:2009du}
\bibitem{Ibanez:2009du} 
  D.~Ibanez, S.~Morisi and J.~W.~F.~Valle,
  ``Inverse tri-bimaximal type-III seesaw and lepton flavor violation'',
  Phys.\ Rev.\ D {\bf 80}, 053015 (2009)
  [arXiv:0907.3109 [hep-ph]]

\bibitem{forero-tortoal-morisi-valle}
 D.~V.~Forero, S.~Morisi, M.~Tortola and J.~W.~F.~Valle,
   ``Lepton flavor violation and non-unitary lepton mixing in low-scale type-I seesaw'',
  JHEP {\bf 1109}, 142 (2011)
  [arXiv:1107.6009 [hep-ph]]

%\cite{Nakamura:2010zzi}
\bibitem{PDG} 
  K.~Nakamura {\it et al.}  [Particle Data Group Collaboration],
  ``Review of particle physics'',
  J.\ Phys.\ G G {\bf 37}, 075021 (2010).
  %%CITATION = JPHGB,G37,075021;%%

\bibitem{meg-futureS}
http://meg.icepp.s.u-tokyo.ac.jp/docs/prop\_psi/proposal.pdf

\bibitem{psi}
{\scriptsize http://www.physi.uni-heidelberg.de/Forschung/he/mu3e/documents/LOI\_Mu3e\_PSI.pdf}

%\cite{Ankenbrandt:2006zu}
%FNAL mu e conversion plan mu2eds
\bibitem{Ankenbrandt:2006zu}
  C.~Ankenbrandt {\it et al.},
  ``Using the Fermilab proton source for a muon to electron conversion
  experiment'',
  arXiv:physics/0611124.
  %%CITATION = PHYSICS/0611124;%%

%\cite{Ilakovac:1994kj}
\bibitem{Ilakovac:1994kj} 
  A.~Ilakovac and A.~Pilaftsis,
  ``Flavor violating charged lepton decays in seesaw-type models'',
  Nucl.\ Phys.\ B {\bf 437}, 491 (1995)
  [hep-ph/9403398].
  %%CITATION = HEP-PH/9403398;%%

%\cite{Adam:2011ch}
\bibitem{Adam:2011ch} 
  J.~Adam {\it et al.}  [MEG Collaboration],
  ``New limit on the lepton-flavour violating decay $\mu^{+} \to e^{+} \gamma$'',
  Phys.\ Rev.\ Lett.\  {\bf 107}, 171801 (2011)
  [arXiv:1107.5547 [hep-ex]].
  %%CITATION = ARXIV:1107.5547;%%

%\cite{Bellgardt:1987du}
\bibitem{Bellgardt:1987du} 
  U.~Bellgardt {\it et al.}  [SINDRUM Collaboration],
  ``Search for the Decay $\mu^+ \to e^+ e^+ e^-$'',
  Nucl.\ Phys.\ B {\bf 299}, 1 (1988).
  %%CITATION = NUPHA,B299,1;%%
\end{thebibliography}


\end{document}